\documentclass[manuscript]{aa}

\usepackage{graphicx}
\usepackage{natbib}
\usepackage{txfonts}
\usepackage{xcolor}
\usepackage{lineno}

\begin{document}

\title{Switchbacks: statistical properties and deviations from alfv\'enicity}

\author{A. Larosa\inst{1}
\and V. Krasnoselskikh\inst{1,2}
\and T. Dudok de Wit\inst{1}
\and O. Agapitov\inst{2}
\and C. Froment\inst{1}
\and V. K. Jagarlamudi\inst{1,11}
\and M. Velli\inst{13}
\and S. D. Bale\inst{2,4,5,6}
\and A. W. Case\inst{7}
\and K. Goetz\inst{8}  
\and P. Harvey\inst{2}
\and J. C. Kasper\inst{12,7,9}
\and K. E. Korreck\inst{7}
\and D. E. Larson\inst{2}
\and R. J. MacDowall\inst{10}
\and D. Malaspina\inst{3,14}
\and  M. Pulupa\inst{2}
\and C. Revillet\inst{1}
\and M. L. Stevens\inst{7}
}

\institute{LPC2E, CNRS/University of Orl\'eans/CNES, 3A avenue de la Recherche Scientifique, Orl\'eans, France\\
\email{andrea.larosa@cnrs-orleans.fr}
\and
Space Sciences Laboratory, University of California, Berkeley, CA 94720-7450, USA
\and
Laboratory for Atmospheric and Space Physics, University of Colorado, Boulder, CO 80303, USA
\and
Physics Department, University of California, Berkeley, CA 94720-7300, USA
\and
The Blackett Laboratory, Imperial College London, London, SW7 2AZ, UK
\and
School of Physics and Astronomy, Queen Mary University of London, London E1 4NS, UK
\and
Smithsonian Astrophysical Observatory, Cambridge, MA, 02138, USA
\and
School of Physics and Astronomy, University of Minnesota, Minneapolis, MN 55455, USA
\and
Climate and Space Sciences and Engineering, University of Michigan, Ann Arbor, MI 48109, USA
\and
Solar System Exploration Division, NASA/Goddard Space Flight Center, Greenbelt, MD, 20771, USA
\and
National Institute for Astrophysics-Institute for Space Astrophysics and Planetology, Via del Fosso del Cavaliere 100, I-00133 Roma, Italy
\and
BWX Technologies, Inc., Washington, DC 20002, USA
\and
Department of Earth, Planetary, and Space Sciences, UCLA, Los Angeles, CA, 90095, USA
\and
Astrophysical and Planetary Sciences Department, University of Colorado, Boulder, CO 80303, USA}

\date{Received; accepted}

\abstract
{Parker Solar Probe's first solar encounter has revealed the presence of sudden magnetic field deflections that are called switchbacks and are associated with proton velocity enhancements in the slow alfv\'{e}nic solar wind.}
{We study their statistical properties with a special focus on their boundaries.}
{Using data from SWEAP and FIELDS we investigate particle and wavefield properties. The magnetic boundaries are analyzed with the minimum variance technique.}
{Switchbacks are found to be alfv\'{e}nic in 73\% of the cases and compressible in 27\%. The correlations between magnetic field magnitude and density fluctuations reveal the existence of both positive and negative correlations, and the absence of perturbations of the magnetic field magnitude. Switchbacks do not lead to a magnetic shear in the ambient field. Their boundaries can be interpreted in terms of rotational or tangential discontinuities. The former are more frequent.}
{Our findings provide constraints on the possible generation mechanisms of switchbacks, which has to be able to account also for structures that are not purely alfv\'{e}nic. One of the possible candidates, among others, manifesting the described characteristics is the firehose instability.} 


\authorrunning{Larosa et al.}
\maketitle

\section{Introduction} 

Two major open questions in Solar Physics are the heating of the solar corona and the acceleration of the solar wind. The Parker Solar Probe (PSP) mission \citep{fox_solar_2016}, which was launched in 2018, offers a unique possibility to shed light on the main questions of Solar Physics: the heating of the solar corona and the acceleration of particles in the solar wind, by making \textit{in situ} measurements in the extended solar corona. One of the most striking results of PSP is the omnipresence of sudden magnetic deflections that have been called jets, switchbacks or velocity spikes \citep{bale_highly_2019,kasper19,ddw20,horbury_sharp_2020}. Switchbacks had been observed before in fast streams of the polar solar wind by the Ulysses spacecraft \citep{Balogh99}. They were recognized as folds in the field, to be distinguished from other \textit{in situ} structures \citep{Yama04}. Switchbacks were also observed in fast wind streams by Helios \citep{Horbury18} and were associated with one-sided radial jets \citep{Gosling2009}.  The first solar encounters of PSP show that they are ubiquitous features of the young slow solar wind. Therefore, their  investigation should help us shed light on their dynamics and evolution in the young solar wind.

In switchbacks, the deflection of the magnetic field occurs simultaneously with that of the proton bulk velocity, which is also enhanced within the structures. 
These structures are considered to be highly alfv\'{e}nic in the sense that the magnitude of the magnetic field hardly varies in time when crossing them and the components of the magnetic field and proton velocity are highly correlated \citep{kasper19}.
Several studies suggest that they are outward propagating alfv\'{e}nic structures \citep{bale_highly_2019, kasper19, horbury_sharp_2020} while others present evidence for long range correlations and indicate that they are rooted deep down in the corona \citep{ddw20}. 
The distribution of the electron pitch angle indicates \citep{whittlesey_solar_2020} that switchbacks are not polarity changes but correspond to reversals of the same magnetic field line.

The radial Poynting flux associated with these structures represents approximately 10\% of the kinetic energy flux
\citep{bale_highly_2019}. However, closer to the Sun, they might carry a more important fraction of energy flux because the ratio of  alfv\'{e}nic flux to proton kinetic flux is inversely proportional to the alfv\'{e}nic Mach number, and so it increases close to the Sun, along with the magnetic field.
The proton radial temperature is higher inside switchbacks \citep{kasper19, Mozer2020, krasnoselskikh_localized_2020}. This correlation seems to be a detailed version of the more general strong correlation found between proton temperature and solar wind speed on larger scales \citep{Grappinetal90,Grappinetal91}, except for structures with 180 degrees rotation \citep{woolley_proton_2020}. 
\citet{woodham_enhanced_nodate} revealed that the parallel temperature is enhanced inside patches of switchbacks, while the perpendicular one is mostly constant, even though higher than the parallel.

Switchback boundaries generally appear as strong discontinuities. The interest discontinuities in the solar wind goes back to the beginning of the space age  \citep{burlaga_micro-scale_1968, Burlaga69}. Such discontinuities are accountable for a significant fraction of the wave power spectral density \citep{borovsky_contribution_2010}. Their type determines the particle and energy exchange between plasma populations inside and outside localized regions such as magnetic field deflections. \citet{phan_parker_2020} showed the absence of reconnection at their boundaries and that the switchbacks boundaries of his data set rather behave as rotational discontinuities. On the contrary \citet{farrell_magnetic_2020} revealed the presence of magnetic dips and \citet{froment_switchbacks_2020} found some cases between 45 to 48 solar radii undergoing reconnection right at the boundaries. This last result is particularly important because reconnection can quickly disintegrate the structures, thereby heating the background solar wind.
\citet{krasnoselskikh_localized_2020}, in case studies, pointed out the presence of strong currents at these boundaries and an enhanced radial Poynting flux. This suggests that some boundary regions may be unstable with respect to the Kelvin-Helmholtz instability \citep{kasper19, Mozer2020}. Finally, these boundaries show enhanced wave activity ranging from MHD scales (as we are going to show) to the whistler frequency range (50-150 Hz) \citep{Agapitov2020, Jagarlamudi_psp_whistlers} and beyond.
There are still many open questions regarding the nature and origin of switchbacks, with a special interest in their role in the solar wind energy balance and plasma heating. Here we address the following questions:
\begin{itemize}
    \item What is the proportion of alfv\'{e}nic versus compressional structures?
    \item What are the sources of the different types of structures and what instabilities could be responsible for their generation?
    \item What are the characteristics of their boundaries and is there any connection between them and the discontinuities that are observed farther away from the Sun (by other missions)? 
\end{itemize}

To answer these questions we hereafter carry out a statistical study of the characteristics of these structures by making use of a manually selected set. 

This paper is organized as follows: in Section~\ref{sec:data} we present the data and the methods; following that we present a statistical analysis of the main properties of switchbacks in Section \ref{sec:properties}, the statistical characteristics of boundaries are presented in Section \ref{section_bound} and summarised in Section \ref{section_summary}. We conclude in Section \ref{section_discuss}.

\section{Data and Methods}
\label{sec:data}

We use the DC magnetic field from the fluxgate magnetometer (MAG) and the AC magnetic field from the search coil magnetometer (SCM); both instruments are part of the FIELDS suite \citep{Bale2016}. The proton density, velocity and temperature are derived from the ion distribution function that is measured by the Faraday from the SWEAP suite \citep{Kasper2016}. The electron density is estimated from the quasi thermal noise technique \citep{moncuquet20}.  We use the RTN coordinate system throughout the paper: the $R$ component is directed anti-sunward along the Sun-spacecraft direction; the tangential $T$ component is the cross product of the solar rotation vector with $R$; the normal $N$ component completes the right-handed set and points in the same direction as the solar rotation vector.

We identified 70 switchbacks with sharp boundaries between 3 November 2018 and 8 November 2018. During this period the cadence of the Faraday cups was of 0.87 seconds while MAG and SCM were sampled either at 146 Hz or at 293 Hz. 
The identification of the switchbacks was made based on the four basic characteristics that are common to all of them: a deflection from the anti-sunward direction of the magnetic field, coming back to the initial conditions, an increase of the magnetic fluctuation amplitude registered by SCM and MAG, and an increase in the proton bulk velocity and radial temperature (this is the only thermal characteristic that is routinely provided by the SWEAP suite \citep{Kasper2016} at the moment).

\begin{figure} 
	\centering
    \includegraphics[width=0.5\textwidth]{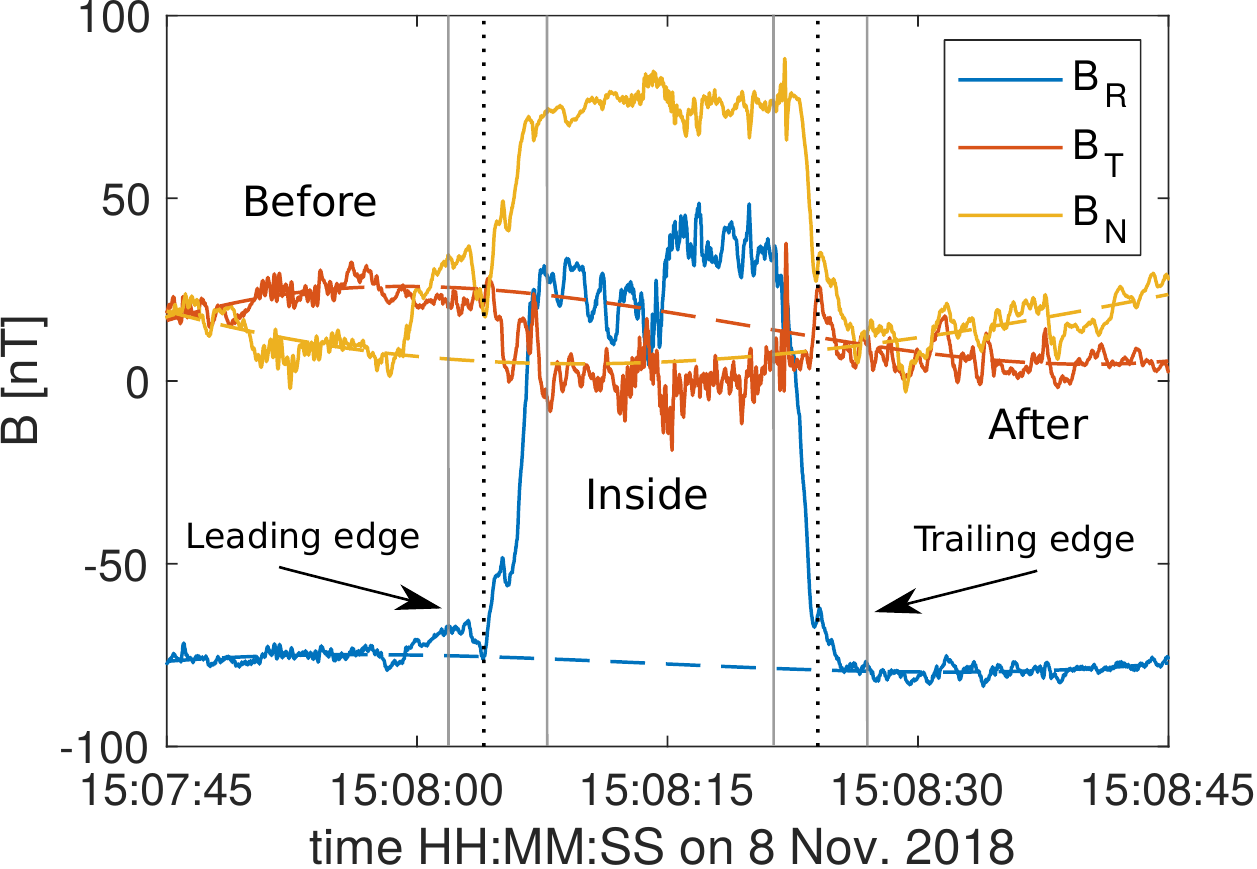} 
    \caption{A typical switchback with a strong deflection and yet well defined edges. Between gray vertical lines the leading and trailing edges. Dashed lines represent the slow evolution of the magnetic field as if no switchback were occurring and are meant to reveal a potential departure of the magnetic field from that trend when approaching the switchback. These trends are obtained from 3rd order polynomial approximation that has been adjusted to the magnetic field up to 5 seconds before the onset of the leading edge of the switchback and starting again 5 seconds after the end of the trailing edge. These transitions are indicated by vertical dotted lines.}
    \label{fig:fig_explicative}
\end{figure}

Figure \ref{fig:fig_explicative} shows a typical switchback signature in the magnetic field from MAG. The boundaries of the structure are located between vertical grey lines. For each structure we record the median magnetic field from MAG and the magnitude of the magnetic field fluctuations from SCM, the velocity, the density and the temperature in the regions indicated in the figure as \emph{Before}, \emph{Inside} and \emph{After} to compare the plasma properties of the structure with respect to the surrounding plasma.  

We study the boundary type, geometry, thickness, current density by applying the minimum variance analysis technique (MVA) \citep{Sonnerup1998} to MAG data. MVA has been extensively used to analyze solar wind discontinuities and current sheets crossings and its limits have been evaluated and validated by comparison with triangulation estimation of the same quantities \citep{horbury_three_2001, knetter_four-point_2004}. In spite of its limitations, MVA is a valuable tool for investigating the geometry and the nature of the discontinuities dealing with single point measurements. Here, for each structure that matches the four above mentioned conditions we apply the technique to its leading and trailing edges.

\begin{figure}
	\centering
    \includegraphics[width=0.5\textwidth]{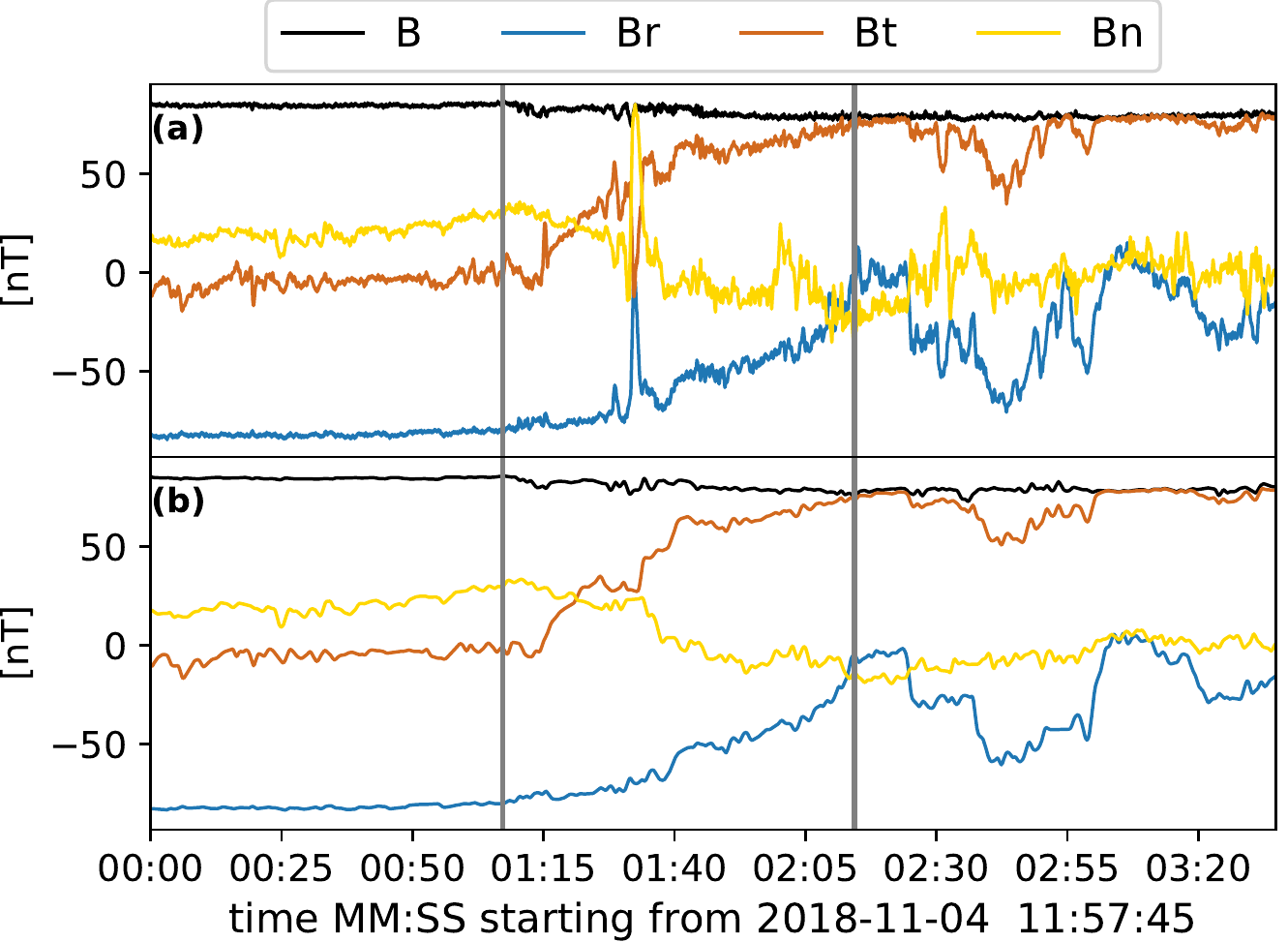} 
    \caption{Example of a spike at the leading edge of a switchback. The upper panel shows the three components and the amplitude of the original MAG data. The lower panel shows the same data after running a median filter with a 10-second window (5 times the duration of the spike) and lowpass filtering. The leading edge of the switchback is located between grey vertical lines.}
    \label{fig:fig_median}
\end{figure}

The application of the MVA is sensitive to the chosen interval around the transition and can be affected by the presence of waves or spike-like structures that are ubiquitous at switchback boundaries. For that reason it is difficult to perform an automated analysis and (subjective) user intervention is often needed. To avoid complex transitions, we only consider cases for which the ratio between intermediate and minimum eigenvalues exceeds 2 \citep{lepping_magnetic_1980} although a ratio higher than 10 has been recommended  \citep{knetter_four-point_2004} to avoid errors due to anisotropic 3D wave activity or surface waves. To further attenuate the impact of wave activity and avoid measuring the wavenumber vector of waves rather than the normal to the boundary (see for example \cite{hudson_discontinuities_1970}), we lowpass filter the magnetic field data, removing frequencies above few Hertz when needed. Another issue is the presence of transient spikes at the edges as already shown by \citet{kasper19}. When present they can mislead the MVA. We remove them with a median filter that is tuned on a case by case basis. Figure~\ref{fig:fig_median} illustrates one example in which the duration of the spike is approximately 2 seconds while the duration of the leading edge is more than one minute.  Here, although the spike is a short-lived transient it may nevertheless alter the result of the MVA analysis. In situations like this we apply the median filter, whose result is shown in the lower panel of Figure~\ref{fig:fig_median}. In the interpretation of the switchbacks as magnetic flux tube crossings \citep{krasnoselskikh_localized_2020} such transients are precisely the signature of wave activity occurring at the surface of the tubes. Here, however, we discard them because our objective is to estimate the normal to the boundary.

\section{Main properties of switchbacks}
\label{sec:properties}
\begin{figure*} 
	\centering
    \includegraphics[width=\textwidth]{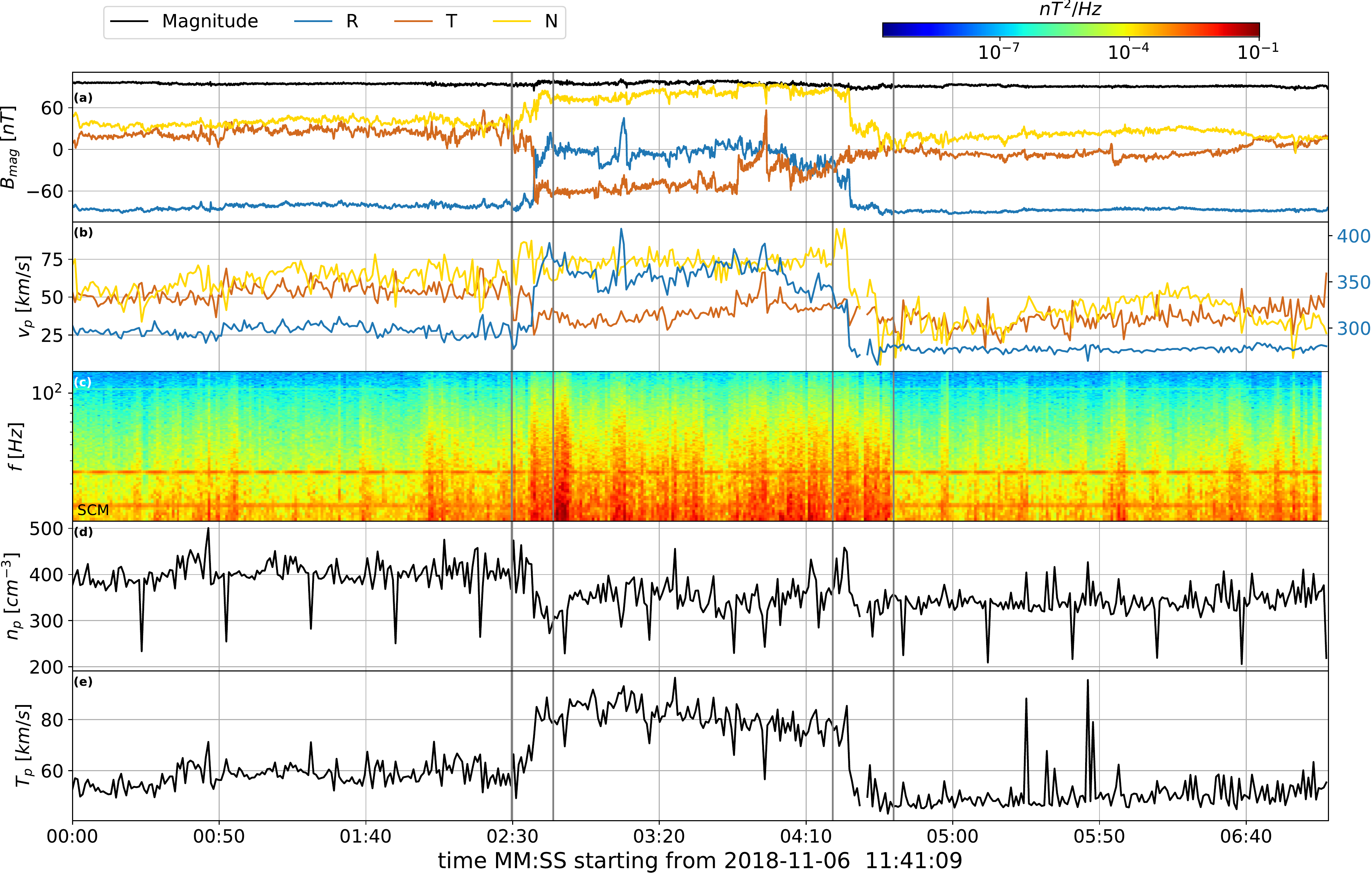} 
    \caption{Switchback example. \textbf{a} MAG magnetic field data, \textbf{b} Proton velocity, \textbf{c} Trace Spectrogram from the SCM waveform, \textbf{d} Proton density, \textbf{e} Proton temperature. Leading and trailing edges, respectively, are located between the first and second pairs of vertical lines.}
    \label{fig:signature}
\end{figure*}

Typical signatures of a switchback are shown in Figure~\ref{fig:signature}. This particular structure has an almost constant magnetic field magnitude, lower proton density inside, increased velocity in both radial and normal components, enhanced wave activity and radial temperature inside. The duration is 130 seconds, while the leading edge lasts 14 seconds and the trailing 20 seconds.

\begin{figure*} 
	\centering
    \includegraphics[width=\textwidth]{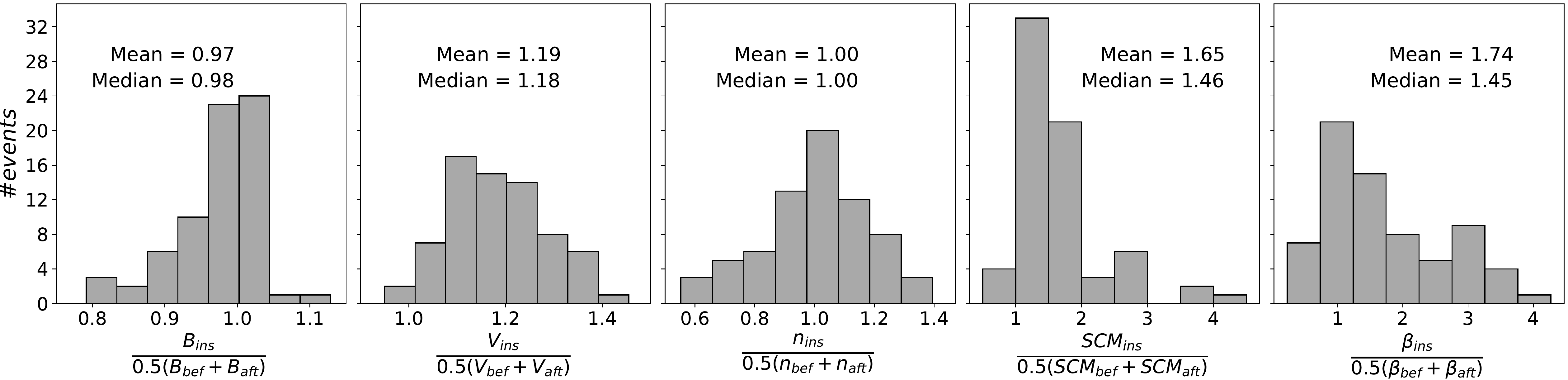} 
    \caption{Main statistical properties of switchbacks, showing for different quantities their average value inside of the switchback relative to that observed outside, see Figure~\ref{fig:fig_explicative}. From left to right: magnetic field magnitude, proton velocity, proton density, amplitude of the magnetic field fluctuations as measured from SCM and radial plasma $\beta$.}
    \label{fig:distri}
\end{figure*}

In Figure \ref{fig:distri} we summarise the main statistical properties of the switchbacks with respect to the surrounding plasma. The first plot represents the distribution of average magnetic field inside relative to the average magnetic field outside. Note the large fraction of alfv\'{e}nic structures, corresponding to the peak at a ratio of 1, which is surrounded by a population of compressible structures. The presence of both types of structures had already been suggested by \citet{bale19, krasnoselskikh_localized_2020}. If we consider as alfv\'{e}nic structures those that are located between 0.95 and 1.05, then 73\% are alfv\'{e}nic and 27\% compressible. 

\begin{figure} 
	\centering
    \includegraphics[width=0.5\textwidth]{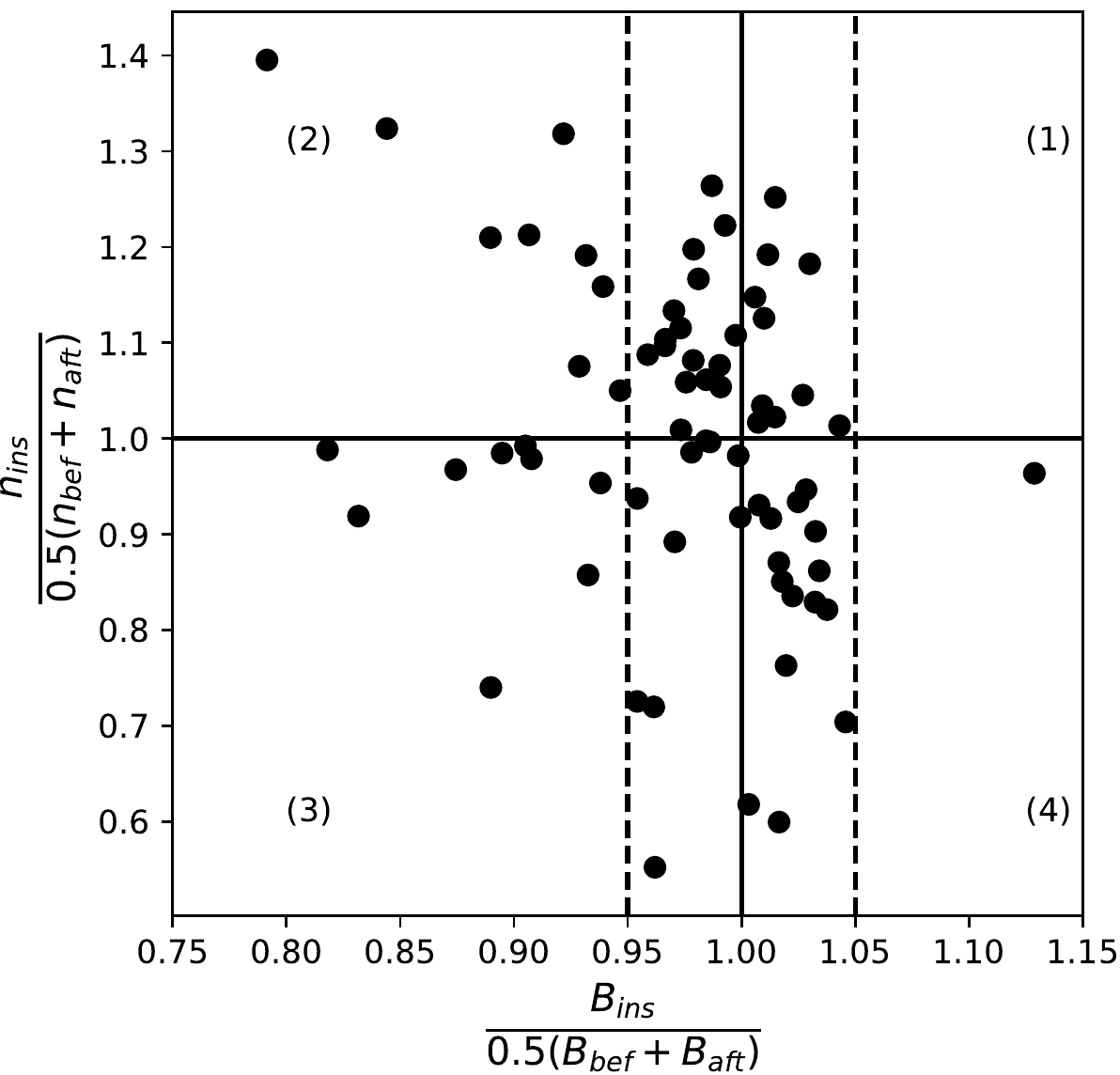} 
    \caption{Comparison of magnetic field and density with ratios of values inside/outside switchbacks. The black vertical and horizontal lines split the values in four quadrants. Quadrants (1) and (3) correspond to structures in which fluctuations are positively correlated, and quadrants (2) and (4) correspond to anticorrelated fluctuations. We consider as alfv\'{e}nic structures those that are located between the two vertical dashed lines.}
    \label{fig:db_dn}
\end{figure}

The second plot illustrates the same ratio of velocities inside/outside and reveals a net increase that is of the order of 100 km/s, which is comparable to the Alfv\'{e}n velocity. The proton density (third plot) and the electron density (not shown) do not show any preferential variation. The magnitude of broadband magnetic fluctuations registered by the SCM (fourth plot) is generally higher inside, except in four cases, with two cases only for which the ratio is below 0.98. The largest relative increases of wave activity are observed for those few  isolated switchbacks that are embedded in a quiet interval with an almost radial field. The plasma radial $\beta$ is systematically higher inside than outside as shown in the fifth plot. When plotted as a function of the deflection angle of the magnetic field across the boundaries this ratio peaks at around 90 degrees. This dependence suggests that higher values of $\beta$ may be caused by a temperature anisotropy rather than by higher temperatures inside the switchbacks. 
On the contrary if we compute the same $\beta$ ratio by artificially keeping a constant temperature across the structures we still have an enhancement of $\beta$ at the interior of the structures. In this case the median value of the ratio decreases from 1.45 to 1.26.
In Figure \ref{fig:db_dn} we plot the magnetic field ratio against the density ratio. Non-alfv\'{e}nic structures lie outside the dashed lines. For ten structures the density and the magnetic field are anticorrelated and for nine they are correlated. This correlation between density and magnetic field variations can be used for wave mode identification. Indeed, in isotropic and homogeneous plasmas such correlations are known to correspond to fast magnetosonic waves when positive whereas anti-correlations correspond to the slow mode \citep{stix1992}. 
This important property was used in the study of the wave activity decomposition on eigenmodes in MHD approximation by \citet{chaston_mhd_2020} making use magnetic field data registered on PSP in the frequency range from 0.0002 to 0.2 Hz during the first encounter. They found that the major wave mode observed belongs to shear Alfv\'{e}n waves, and there is about 20-30 \% of energy carried by the slow mode waves. It is worth noting that the authors excluded switchback boundaries form their analysis and so could not identify which wave-mode should be attributed the switchbacks themselves.

The correlation between density and magnetic variations does not show any dependence on deflection angle of the switchback. Note that the less structured and shorter-lived switchbacks are mostly alfv\'{e}nic. 

Another question of interest is whether switchbacks modify the magnetic field in their vicinity, for example because of currents flowing in or along it. Quantifying such a distortion is challenging because it requires switchbacks with sharp edges that are surrounded by a quiet magnetic field without trends or evidence for other or embedded switchbacks. We found only a few tens of such cases during the first encounter of PSP. Figure~\ref{fig:fig_explicative} illustrates one of them, which in addition has a strong deflection of more than $90^{\circ}$. To determine a possible distortion we extrapolate here the magnetic field during the switchback by considering its value up to 5 seconds before the onset of the leading edge and again 5 seconds after the end of the trailing edge. The figure shows scarce evidence for a distortion of the magnetic field near the edges. In a total of 16 switchbacks we found with sharp edges and a reversal of more than 90 degrees, only two showed weak evidence for a distortion. We conclude that the current flowing at the boundary of switchbacks does not affect the surrounding magnetic field in a significant way.

\subsection{Size and currents}
\label{sec:size_currents}

\begin{figure}
	\centering
    \includegraphics[width=0.5\textwidth]{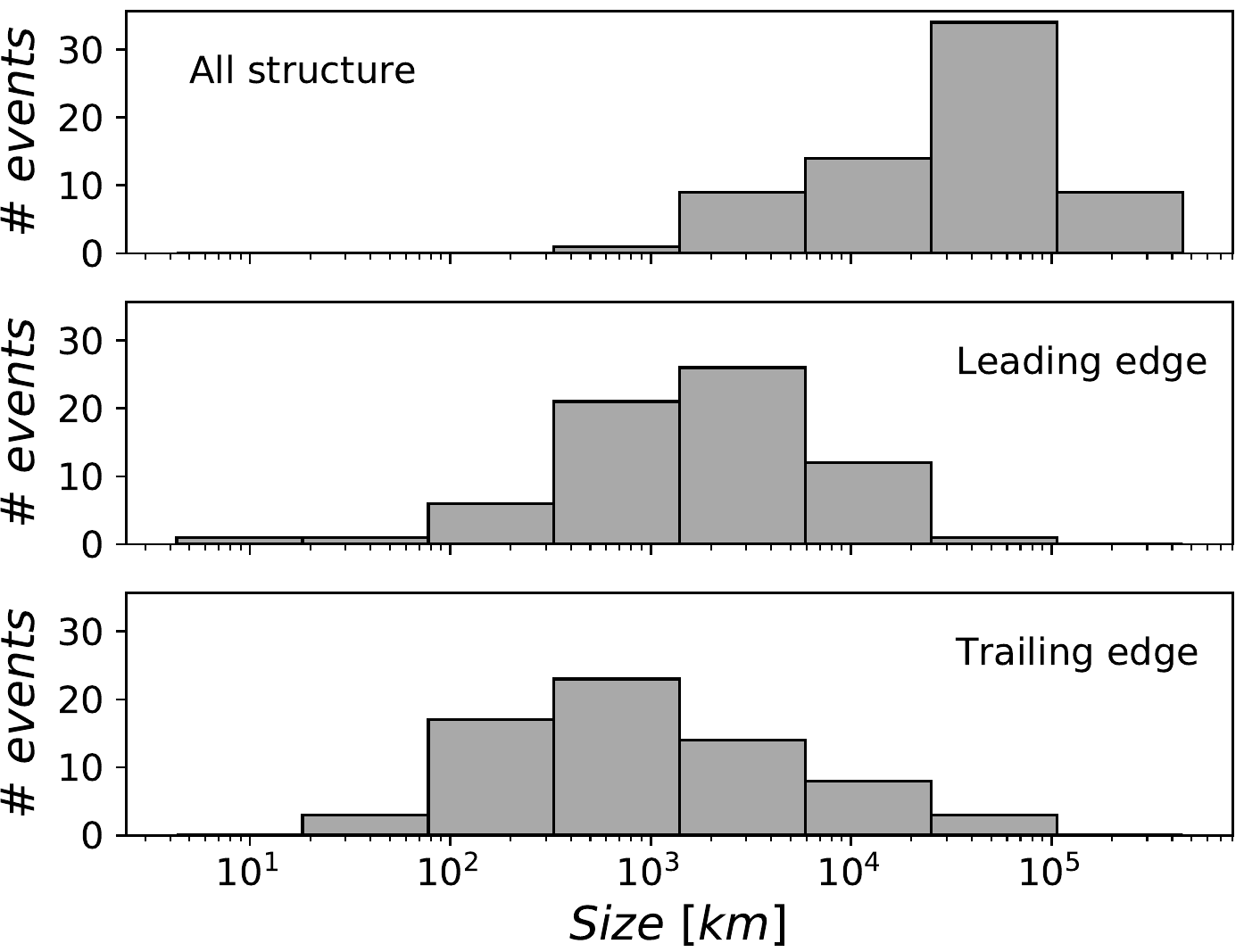} 
    \caption{Distribution of the sizes of the switchbacks edges and full structure.}
    \label{fig:size}
\end{figure}

The transverse size of switchbacks can be estimated from the product of the normal velocity (the projection of the proton velocity along the normal direction $\hat{n} $) multiplied by the crossing time. Since we have two different normal velocities at leading and trailing edges we first compute the size of both edges. We evaluate the size of the whole structure as the sum of the two edges plus the size of the internal region. To estimate the latter, we use the average of the velocities at the leading and trailing edges multiplied by the crossing time. The results are shown in Figure~\ref{fig:size}. The size of the largest structure in our data set is approximately 0.65 $R_{sun}$.

The sizes of the boundaries exceed the ion inertial length between one and three orders of magnitude and the structures may therefore be interpreted in terms of MHD discontinuities.

\begin{figure}
	\centering
    \includegraphics[width=0.5\textwidth]{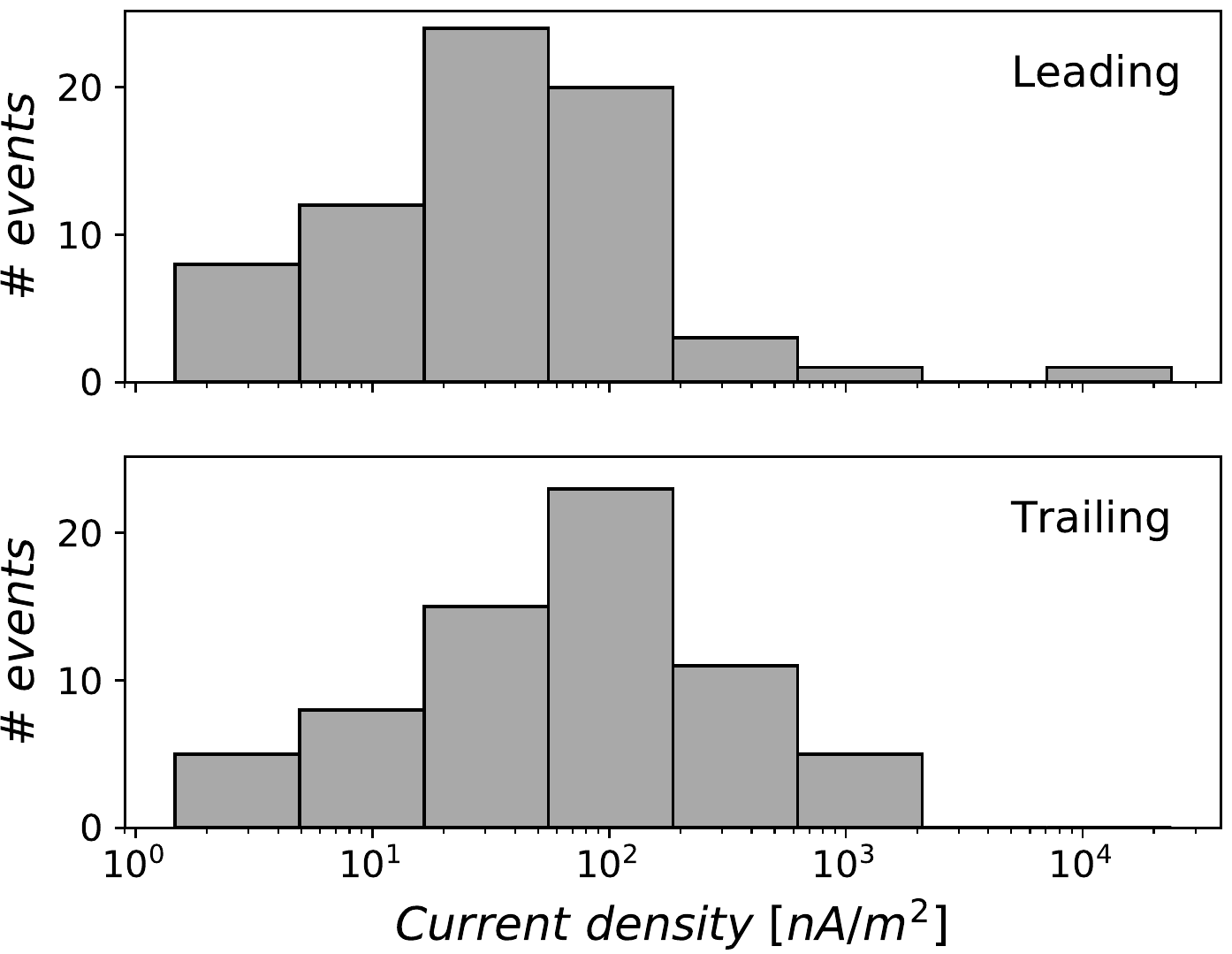} 
    \caption{Distribution of the magnitude of the current density at switchback boundaries.}
    \label{fig:current}
\end{figure}

Making use the technique presented by \citet{krasnoselskikh_localized_2020} we estimated the currents from the rotation of the magnetic field between both sides of each boundary. At MHD scales we have $ \vec{J} = \vec{\nabla} \times \vec{B}/ \mu_0 $. Moving across a switchback boundary the spatial variation happens along the minimum variance direction, while the component of the field that has a considerable gradient is the maximum variance component. Therefore, the cross product of the two that is along the intermediate variance direction should give an estimate of the current. A similar approach had also been used in \citep{artemyev_dynamics_2018}. The result of this procedure are shown in Figure~\ref{fig:current}.
The currents we find are between one and two orders of magnitude larger than those observed in solar wind discontinuities at 1 AU  \citep{artemyev_dynamics_2018}, even though the sizes of the discontinuities are comparable. The main difference between the two resides in the stronger magnetic field at the location of PSP.

\section{Boundaries}
\label{section_bound}

\begin{figure} 
	\centering
    \includegraphics[width=0.45\textwidth]{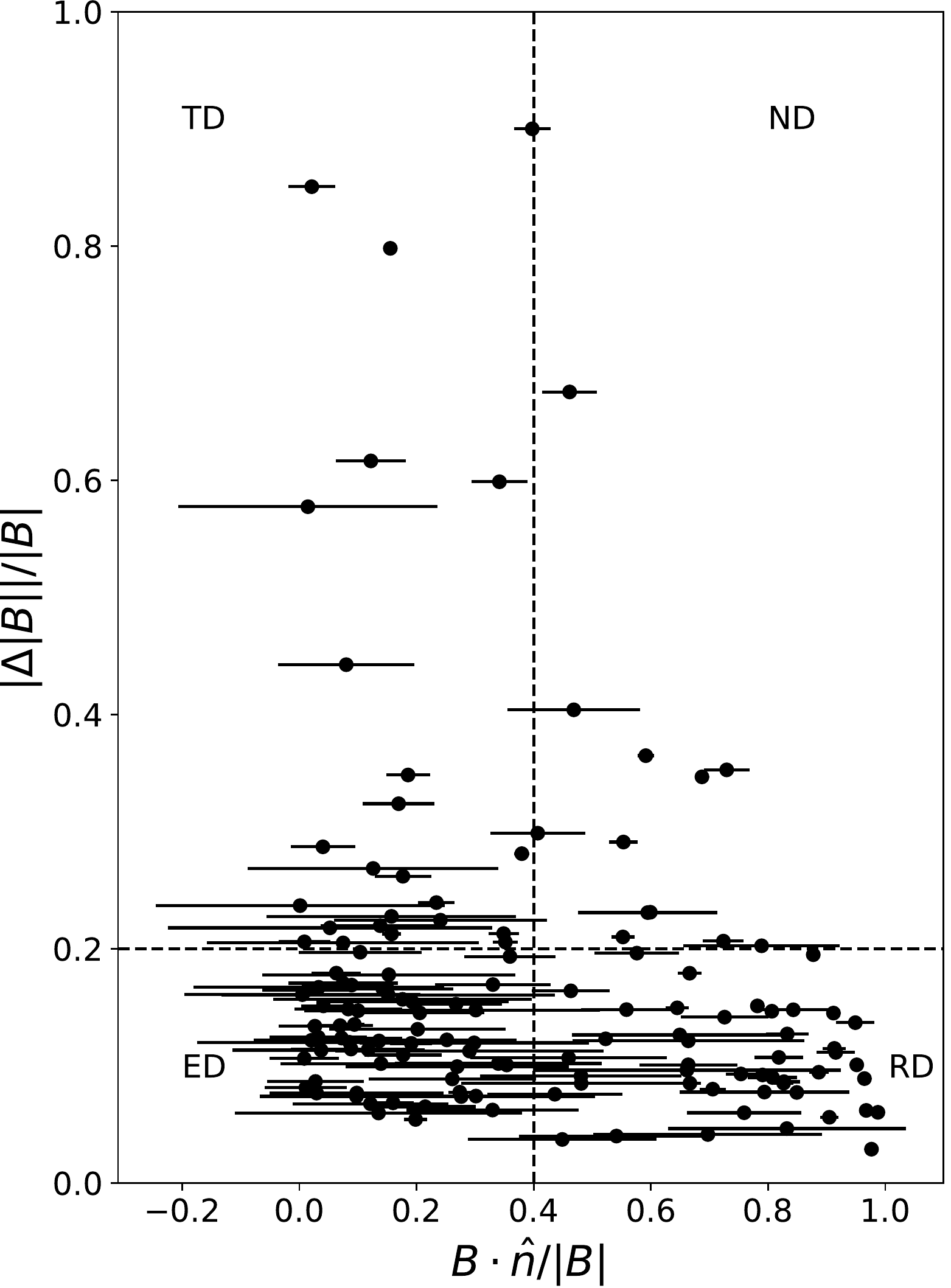} 
    \caption{Boundary classification comparing the field threading through the discontinuity surface normalized to the field magnitude ($x$ axis) and the highest median magnetic field magnitude between the two sides of the discontinuity ($y$ axis). The dotted lines separate different types of discontinuities. The $\pm 1 \sigma$ error bars in the $x$ direction are estimated by block bootstrapping.}
    \label{fig:rot_tan}
\end{figure}

The results of our MVA analysis of switchback boundaries are summarised in Figure~\ref{fig:rot_tan}. Here we use the same classification scheme as  \citet{neugebauer_reexamination_1984} and \citet{horbury_three_2001} for both leading and trailing edges combined together, in order to have a better statistics. 

Discontinuities with $ \vec{B} \cdot \hat{n} / |B| >0.4$ and $ \delta |B|/|B|<0.2$ are supposedly rotational discontinuities (RDs) because they have a significant field component normal to the surface (considered as plane) while the magnitude of the variation is still consistent with alfv\'{e}nic fluctuations. Discontinuities with $ \vec{B} \cdot \hat{n} / |B| <0.4$ and $ \delta |B|/|B|>0.2$ are classified as tangential discontinuities (TDs) because there of the major change in magnitude while the field threading is not a large fraction of the total field. 
The area notified by 'Either' (ED) corresponds to values of parameters for which it is difficult to unambiguously distinguish between rotational and tangential discontinuities; the area marked as 'Neither' (ND) is  inconsistent with characteristics of the MHD discontinuities and indeed is mostly empty.

We found that 32\% of the switchbacks boundaries are RDs, 17\% are TDs, 42\% are in the 'Either' area and 9\% are in the 'Neither' area. These results are consistent with similar studies performed at 1 AU (see Table 2 of \citep{neugebauer_comment_2006}). The higher occurrence of RDs with respect to TDs is expected for alfv\'{e}nic winds emerging from coronal holes \citep{soding_radial_2001}.

Note that periods when $ \delta |B|/|B|$ is above 40\% are typically due to sharp dips in the magnetic field that happen right at the switchback boundaries. The field quickly recovers its magnitude once the boundary is crossed. This is the reason why such sharp changes are not observed in the first distribution of figure \ref{fig:distri}.
The structures for which $ \vec{B} \cdot \hat{n} / |B|$  is close to one are switchbacks in which the MVA direction is close to the radial direction. For that reason they have large normal component. This happens for alfv\'{e}nic fluctuations that propagate nearly parallel to the ambient magnetic field \citep{soding_radial_2001}.
TDs can be generated either at the stream interface, by the mirror instability \citep{soding_radial_2001} or by the diamagnetic component of boundaries current \citep{krasnoselskikh_localized_2020, farrell_magnetic_2020}.

This gives two scenarios for switchbacks that are embedded between two TDs. In the first one, since no significant velocity shear is observed across the boundaries, the switchback has to be generated deeper down in the corona where supposedly the shear occurred. In the second case, the TDs would be just the result of the mirror instability at switchbacks boundaries or the effect of the diamagnetic currents.

RDs could be the steepened edge of a large amplitude Alfv\'{e}n waves \citep{tsurutani_relationship_1994}, a picture that is consistent with the constant magnetic field magnitude switchbacks with rotational discontinuities as boundaries. However, in most cases we find the type of the discontinuity to be different at the leading and trailing edges. Most switchbacks have a TD at their leading edge and RD at their trailing edge, or vice-versa.

\subsection{Wave activity at boundaries}
\label{wav_boundary}

\begin{figure} 
	\centering
    \includegraphics[width=0.4\textwidth]{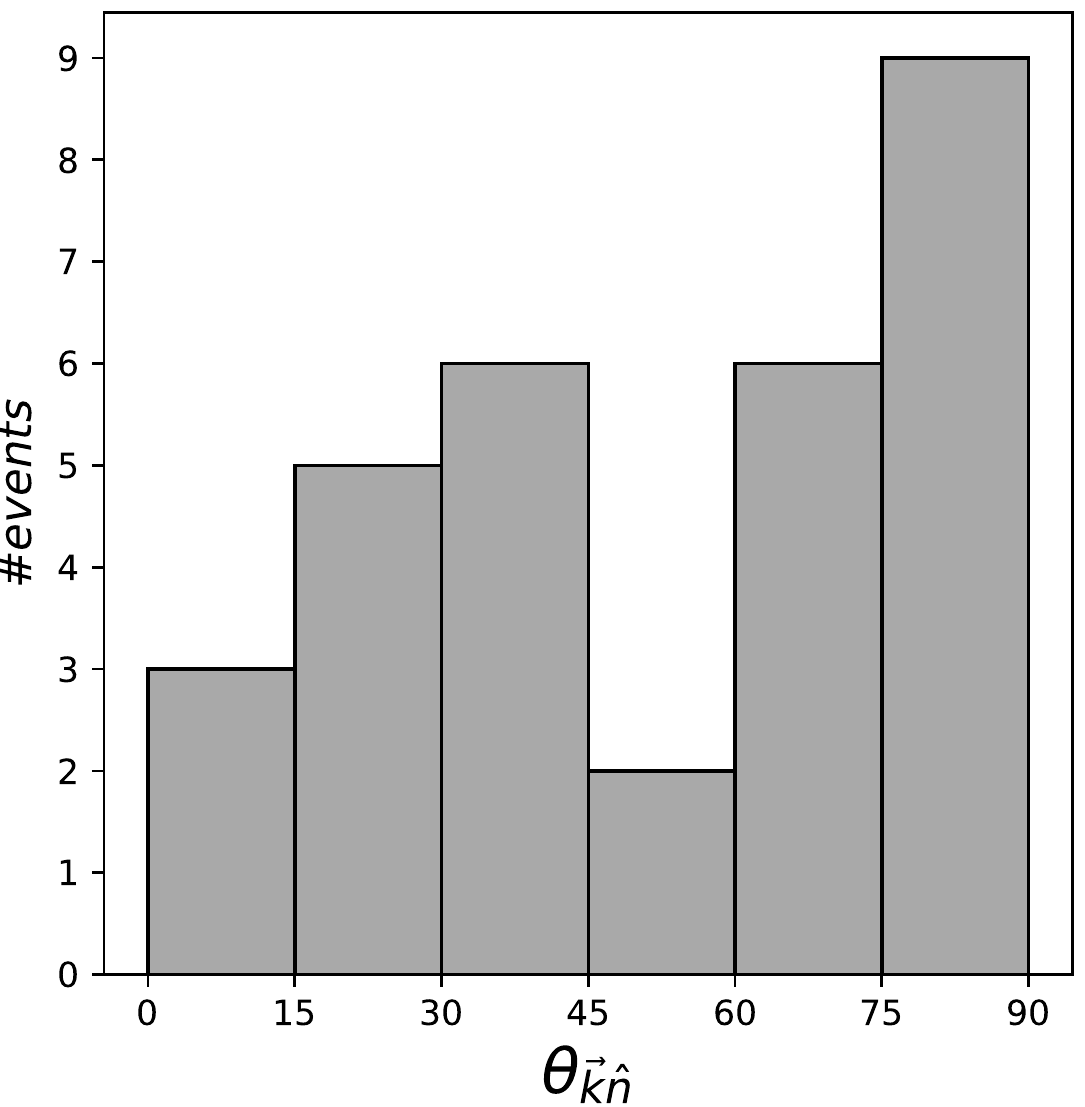} 
    \caption{Distribution of the angles between the normal to the boundary and the $k$ vector of the wave located at that boundary.}
    \label{fig:small_hist}
\end{figure}

The MAG and SCM magnetometers often reveal enhanced levels of magnetic wave activity at switchback boundaries. Most waves are broadband with frequencies typically ranging between 0.1 and 30 Hz. To analyse these waves we bandpass filter them and perform a polarisation analysis \citep{santolik_singular_2003} in the frequency band that has the largest power. This analysis is complemented by visual inspection of the hodograms. In the following we only consider those wave whose planarity exceeds 0.7.
Of the 140 switchback boundaries we analysed only 32 present revealed polarised waves matching our criteria. Most of these waves have frequencies around 3 Hz. The angle between the normal to the boundary and their wavenumber vector (see Figure~\ref{fig:small_hist}) is restricted between 0 and 90 degrees, due to the intrinsic ambiguity of the polarisation analysis method. With this, we find the median value to be approximately 60 degrees; this large angle is consistent with the presence of surface waves as suggested by \citet{krasnoselskikh_localized_2020}. Waves propagating more parallel to the boundary normal are probably gradient drift waves that are generated by the gradients of the plasma parameters across switchback boundaries.

\section{Summary of observations}
\label{section_summary}

We have presented a statistical study of the characteristics of the switchbacks and their boundaries. Their characteristic transverse sizes vary from several thousands of kilometers up to the solar radius. Some of them, mainly small scale, are quite uniform, others have well pronounced internal substructures. We showed the presence of two different types of switchbacks, alfv\'{e}nic and compressible. The former are more frequent than the second. 
Our study also reveals that the majority of switchbacks present important plasma density variations from outside to inside. The density can either increase or decrease but typically does not remain the same as outside. We found that the density and magnetic field jumps are positively correlated for some switchbacks and anti-correlated for others, but most of them present density jumps even when the magnetic field strength remains practically constant. 
This density jumps are observed with both SPAN-i and SPC.
The wave activity is enhanced for the majority of switchback at their interior and for few of them at their boundaries, where surface waves are probably present. The radial $\beta$ is enhanced inside the structures.
The investigation of switchback boundaries led us to their interpretation in terms of MHD discontinuities since their size is orders of magnitude larger than the ion inertial length and ion Larmor radius. The classification of the boundaries shows the presence of both rotational and tangential discontinuities, the former being twice more frequent. Strong currents occur at the boundaries of switchbacks, this corroborates the idea proposed by \citet{krasnoselskikh_localized_2020} that these structures are localized kinked magnetic tubes separated from surrounding plasma by surface currents. 

\section{Discussion and Conclusions}
\label{section_discuss}

Our statistical study aims at providing possible constraints imposed by the characteristics of switchbacks, which may in turn help determine the possible generation mechanisms of these structures and their possible role in heating and accelerating the solar wind. The large variety of scales and characteristics suggests that several mechanisms generate switchbacks. To the best of our knowledge four mechanisms have been proposed so far.

The first mechanism was originally proposed to explain observations made by Ulysses of so-called interchange reconnection between closed and open magnetic field lines \citet{Yamab04}. More recently this explanation has been invoked by \citet{fisk_global_2020} to explain the generation of  switchbacks at the boundary of the small coronal holes due to reconnection between open and closed field lines. This transient reconnection supposedly occurs in low-$\beta$ plasmas. Following this, alfv\'{e}nic and slow mode type perturbations propagate along open field lines, generating a helicity that is eventually evacuated by means of convection to interplanetary space in the form of alfv\'{e}nic-type perturbations  \citep[see][]{edmondson_role_2012}. Such processes may naturally occur inside and at the boundary of coronal holes \citep{Yamab04, fisk_acceleration_2003}. Numerous switchbacks were indeed observed PSP during its first solar encounters whenever the spacecraft was connected to small coronal holes \citep{Badman2020, Panasenco20}. In contrast, were very few switchbacks occurred during the second  encounter, when there was no such connection. The characteristics of structures we identified are consistent with such a mechanism. In particular, interchange reconnection may explain the alfv\'{e}nic and compressible structures that correspond to the anti-correlation of density and magnetic field variations. Fast mode like structures are unlikely to arise with interchange reconnection \citep{kigure_generation_2010}.

A second plausible mechanism arises from the evolution of Alfv\'en waves propagating in the expanding solar wind \citep{Hollweg74, Velli93, Landietal06}. In the presence of radial gradients, the amplitude of fluctuations grows relative to the mean field, so that  spherically polarized  Alfv\'{e}n waves with a small radial component may grow until a switchback is formed. This mechanism was demonstrated by means of computer simulations based on the expanding box model \citep{velli92,grappin96,squire_situ_2020}. It gives rise to a predominant population of alfv\'{e}nic modes, with small slow or fast contributions depending on the plasma $\beta$. However, we are lacking information about the range of sizes of the structures that may be generated that way.  

A third possibility is the interaction of shear flows, which are naturally present in the young solar wind, with the omnipresent fluctuations that propagate away from the Sun. This mechanism was first proposed by \citet{Landietal05} and \citet{Landietal06}. A slightly different version, involving the nonlinear instability of such sheared flows, has been proposed by Ruffolo et al. 2020 (ApJ, in press, private communication). This mechanism mainly leads to a slow-mode (pressure-balanced) contribution to the switchback. Such structures exhibit an anticorrelation between the density and magnetic field variations, in agreement with our observations. However, they are expected to be less frequent than alfv\'enic structures.

Finally, a fourth mechanism is the generation of perpendicular magnetic perturbations due to microscopic plasma instabilities. Indeed, the excess of parallel streaming due to either pressure anisotropy or to the presence of a secondary beam can lead to structures that are similar to switchbacks. \citet{tenerani_nonlinear_2018}, using simulations, have shown how magnetic field reversals bounded by rotational discontinuities can be a natural consequence of the nonlinear evolution of right handed circularly polarized Alfv\'{e}n waves in high-$\beta$ plasma regimes that are unstable with respect to the firehose instability. The proton temperature anisotropy measured by PSP is not favorable to trigger such an instability as proton beams are mostly observed in quiet radial field regions \citep{verniero_parker_2020}. Nevertheless, remote sensing observations show the presence of numerous jets in the lower corona \citep{sterling20}. The velocity of these jets may exceed the Alfv\'{e}n velocity, thus offering favorable conditions for the firehose instability. 

Another possible source of instability of right handed polarized Alfv\'en waves may be related to the presence of alpha particles. These particles show a tendency to have $T_{\parallel} > T_{\perp}$ close to the Sun  \citep{stansby_alpha_2019} and numerical simulations have shown that they can trigger the firehose instability \citep{matteini_fire_2015}. Note that a firehose-like mode could also grow by cyclotron resonant interaction of a parallel propagating wave of the fast magnetosonic/whistler branch  with alpha particles in presence of an alpha-proton drift equal or greater than  1.7 Alfv\'{e}n speeds
 \citep{verscharen_dispersion_2013}. 
The oblique firehose instability initially gives rise to non propagating modes with anti-correlated magnetic and density fluctuations; these modes eventually turn into elliptically polarized Alfv\'en waves  \citep{hellinger_new_2000}. This mechanism could produce the slow mode-like anti-correlation observed for some of the  switchbacks.    
It is worth noting that the microscopic plasma instabilities may account for small and moderate scale perturbations, though it is difficult to determine their limiting scale because of the effect of the solar wind expansion.


What key features could help determine the most plausible generation mechanisms ? In this regard the presence of alfv\'{e}nic, fast and slow mode signatures in switchbacks is important. There is a caveat, however. Our analysis does not unambiguously attribute the structures to a certain type of wave-mode, since it is based only on the correlation and anti-correlation of the density perturbations and magnetic field magnitude. 

As shown by \citet{hau_slow_mode_1993}, in anisotropic plasmas, while fast modes always show a positive correlation, slow modes can lead to both positive and negative correlations depending on parameters such as the angle between the wavenumber vector and the magnetic field, level of anisotropy, and chosen state equation for moments closure. Of particular interest is their result showing that under anisotropic conditions, favourable to the firehose instability, slow modes exhibit a positive correlation for a wider range of angles between magnetic field and wavenumber.

From this we conclude that the only two mechanisms that are able to explain all three types of observed modes are the evolution of Alfv\'{e}n waves in the expanding solar wind and firehose-like instabilities.

\begin{acknowledgements}
The FIELDS experiment was developed and is operated under NASA contract NNN06AA01C. AL, VK, TD, CF, VKJ and CR acknowledge financial support of CNES in the frame of Parker Solar Probe grant. SDB acknowledges the support of the Leverhulme Trust Visiting Professorship programme. O.A. and V.K. were supported by NASA grant 80NSSC20K0697. O.A. was partially supported by NASA grants 80NNSC19K0848, 80NSSC20K0218, and NSF grant NSF 1914670. Parker Solar Probe was designed, built, and is now operated by the Johns Hopkins Applied Physics Laboratory as part of NASA’s Living with a Star (LWS) program (contract NNN06AA01C). Support from the LWS management and technical team has played a critical role in the success of the Parker Solar Probe mission. The data used in this study are available at the NASA Space Physics Data Facility (SPDF), \url{https://spdf.gsfc.nasa.gov}.
\end{acknowledgements}

\bibliographystyle{aa.bst}                        
\bibliography{paper_mio}

\begin{thebibliography}{63}
\expandafter\ifx\csname natexlab\endcsname\relax\def\natexlab#1{#1}\fi

\bibitem[{{Agapitov} {et~al.}(2020){Agapitov}, {Dudok de Wit}, {Mozer},
  {Bonnell}, {Drake}, {Malaspina}, {Krasnoselskikh}, {Bale}, {Whittlesey},
  {Case}, {Chaston}, {Froment}, {Goetz}, {Goodrich}, {Harvey}, {Kasper},
  {Korreck}, {Larson}, {Livi}, {MacDowall}, {Pulupa}, {Revillet}, {Stevens}, \&
  {Wygant}}]{Agapitov2020}
{Agapitov}, O.~V., {Dudok de Wit}, T., {Mozer}, F.~S., {et~al.} 2020, The
  Astrophysical Journal Letters, accepted

\bibitem[{Artemyev {et~al.}(2018)Artemyev, Angelopoulos, Halekas, Vinogradov,
  Vasko, \& Zelenyi}]{artemyev_dynamics_2018}
Artemyev, A.~V., Angelopoulos, V., Halekas, J.~S., {et~al.} 2018, The
  Astrophysical Journal, 859, 95

\bibitem[{{Badman} {et~al.}(2020){Badman}, {Bale}, {Mart{\'\i}nez Oliveros},
  {Panasenco}, {Velli}, {Stansby}, {Buitrago-Casas}, {R{\'e}ville}, {Bonnell},
  {Case}, {Dudok de Wit}, {Goetz}, {Harvey}, {Kasper}, {Korreck}, {Larson},
  {Livi}, {MacDowall}, {Malaspina}, {Pulupa}, {Stevens}, \&
  {Whittlesey}}]{Badman2020}
{Badman}, S.~T., {Bale}, S.~D., {Mart{\'\i}nez Oliveros}, J.~C., {et~al.} 2020,
  \apjs, 246, 23

\bibitem[{Bale {et~al.}(2019)Bale, Badman, Bonnell, Bowen, Burgess, Case,
  Cattell, Chandran, Chaston, Chen, Drake, de~Wit, Eastwood, Ergun, Farrell,
  Fong, Goetz, Goldstein, Goodrich, Harvey, Horbury, Howes, Kasper, Kellogg,
  Klimchuk, Korreck, Krasnoselskikh, Krucker, Laker, Larson, MacDowall,
  Maksimovic, Malaspina, Martinez-Oliveros, McComas, Meyer-Vernet, Moncuquet,
  Mozer, Phan, Pulupa, Raouafi, Salem, Stansby, Stevens, Szabo, Velli, Woolley,
  \& Wygant}]{bale_highly_2019}
Bale, S.~D., Badman, S.~T., Bonnell, J.~W., {et~al.} 2019, Nature, 576, 237

\bibitem[{{Bale} {et~al.}(2019){Bale}, {Badman}, {Bonnell}, {Bowen}, {Burgess},
  {Case}, {Cattell}, {Chandran}, {Chaston}, {Chen}, {Drake}, {de Wit},
  {Eastwood}, {Ergun}, {Farrell}, {Fong}, {Goetz}, {Goldstein}, {Goodrich},
  {Harvey}, {Horbury}, {Howes}, {Kasper}, {Kellogg}, {Klimchuk}, {Korreck},
  {Krasnoselskikh}, {Krucker}, {Laker}, {Larson}, {MacDowall}, {Maksimovic},
  {Malaspina}, {Martinez-Oliveros}, {McComas}, {Meyer-Vernet}, {Moncuquet},
  {Mozer}, {Phan}, {Pulupa}, {Raouafi}, {Salem}, {Stansby}, {Stevens}, {Szabo},
  {Velli}, {Woolley}, \& {Wygant}}]{bale19}
{Bale}, S.~D., {Badman}, S.~T., {Bonnell}, J.~W., {et~al.} 2019, \nat, 576, 237

\bibitem[{{Bale} {et~al.}(2016){Bale}, {Goetz}, {Harvey}, {Turin}, {Bonnell},
  {Dudok de Wit}, {Ergun}, {MacDowall}, {Pulupa}, {Andre}, {Bolton},
  {Bougeret}, {Bowen}, {Burgess}, {Cattell}, {Chandran}, {Chaston}, {Chen},
  {Choi}, {Connerney}, {Cranmer}, {Diaz-Aguado}, {Donakowski}, {Drake},
  {Farrell}, {Fergeau}, {Fermin}, {Fischer}, {Fox}, {Glaser}, {Goldstein},
  {Gordon}, {Hanson}, {Harris}, {Hayes}, {Hinze}, {Hollweg}, {Horbury},
  {Howard}, {Hoxie}, {Jannet}, {Karlsson}, {Kasper}, {Kellogg}, {Kien},
  {Klimchuk}, {Krasnoselskikh}, {Krucker}, {Lynch}, {Maksimovic}, {Malaspina},
  {Marker}, {Martin}, {Martinez-Oliveros}, {McCauley}, {McComas}, {McDonald},
  {Meyer-Vernet}, {Moncuquet}, {Monson}, {Mozer}, {Murphy}, {Odom},
  {Oliverson}, {Olson}, {Parker}, {Pankow}, {Phan}, {Quataert}, {Quinn},
  {Ruplin}, {Salem}, {Seitz}, {Sheppard}, {Siy}, {Stevens}, {Summers}, {Szabo},
  {Timofeeva}, {Vaivads}, {Velli}, {Yehle}, {Werthimer}, \&
  {Wygant}}]{Bale2016}
{Bale}, S.~D., {Goetz}, K., {Harvey}, P.~R., {et~al.} 2016, \ssr, 204, 49

\bibitem[{{Balogh} {et~al.}(1999){Balogh}, {Forsyth}, {Lucek}, {Horbury}, \&
  {Smith}}]{Balogh99}
{Balogh}, A., {Forsyth}, R.~J., {Lucek}, E.~A., {Horbury}, T.~S., \& {Smith},
  E.~J. 1999, \grl, 26, 631

\bibitem[{Borovsky(2010)}]{borovsky_contribution_2010}
Borovsky, J.~E. 2010, Physical Review Letters, 105, 111102

\bibitem[{Burlaga(1968)}]{burlaga_micro-scale_1968}
Burlaga, L.~F. 1968, Solar Physics, 4, 67

\bibitem[{{Burlaga} \& {Ness}(1969)}]{Burlaga69}
{Burlaga}, L.~F. \& {Ness}, N.~F. 1969, \solphys, 9, 467

\bibitem[{Chaston {et~al.}(2020)Chaston, Bonnell, Bale, Kasper, Pulupa, Wit,
  Bowen, Larson, Whittlesey, Wygant, Salem, MacDowall, Livi, Vech, Case,
  Stevens, Korreck, Goetz, Harvey, \& Malaspina}]{chaston_mhd_2020}
Chaston, C.~C., Bonnell, J.~W., Bale, S.~D., {et~al.} 2020, The Astrophysical
  Journal Supplement Series, 246, 71

\bibitem[{{Dudok de Wit} {et~al.}(2020){Dudok de Wit}, {Krasnoselskikh},
  {Bale}, {Bonnell}, {Bowen}, {Chen}, {Froment}, {Goetz}, {Harvey},
  {Jagarlamudi}, {Larosa}, {MacDowall}, {Malaspina}, {Matthaeus}, {Pulupa},
  {Velli}, \& {Whittlesey}}]{ddw20}
{Dudok de Wit}, T., {Krasnoselskikh}, V.~V., {Bale}, S.~D., {et~al.} 2020,
  \apjs, 246, 39

\bibitem[{Edmondson(2012)}]{edmondson_role_2012}
Edmondson, J.~K. 2012, Space Science Reviews, 172, 209

\bibitem[{Farrell(2020)}]{farrell_magnetic_2020}
Farrell, W.~M. 2020, The Astrophysical Journal Supplement Series, 10

\bibitem[{Fisk(2003)}]{fisk_acceleration_2003}
Fisk, L.~A. 2003, Journal of Geophysical Research, 108, 1157

\bibitem[{Fisk \& Kasper(2020)}]{fisk_global_2020}
Fisk, L.~A. \& Kasper, J.~C. 2020, The Astrophysical Journal, 894, L4

\bibitem[{Fox {et~al.}(2016)Fox, Velli, Bale, Decker, Driesman, Howard, Kasper,
  Kinnison, Kusterer, Lario, Lockwood, McComas, Raouafi, \&
  Szabo}]{fox_solar_2016}
Fox, N.~J., Velli, M.~C., Bale, S.~D., {et~al.} 2016, Space Science Reviews,
  204, 7

\bibitem[{Froment {et~al.}(2021)Froment, Krasnoselskikh, Dudok~de Wit,
  Agapitov, Fargette, Lavraud, Kretzschmar, Jagarlamudi, \&
  Larosa}]{froment_switchbacks_2020}
Froment, C., Krasnoselskikh, V., Dudok~de Wit, T., {et~al.} 2021, under review,
  Astronomy and Astrophysics, this issue

\bibitem[{{Gosling} {et~al.}(2009){Gosling}, {McComas}, {Roberts}, \&
  {Skoug}}]{Gosling2009}
{Gosling}, J.~T., {McComas}, D.~J., {Roberts}, D.~A., \& {Skoug}, R.~M. 2009,
  \apjl, 695, L213

\bibitem[{{Grappin} {et~al.}(1990){Grappin}, {Mangeney}, \&
  {Marsch}}]{Grappinetal90}
{Grappin}, R., {Mangeney}, A., \& {Marsch}, E. 1990, \jgr, 95, 8197

\bibitem[{Grappin \& Velli(1996)}]{grappin96}
Grappin, R. \& Velli, M. 1996, Journal of Geophysical Research: Space Physics,
  101, 425

\bibitem[{{Grappin} {et~al.}(1991){Grappin}, {Velli}, \&
  {Mangeney}}]{Grappinetal91}
{Grappin}, R., {Velli}, M., \& {Mangeney}, A. 1991, Annales Geophysicae, 9, 416

\bibitem[{Hau \& Sonnerup(1993)}]{hau_slow_mode_1993}
Hau, L.~N. \& Sonnerup, B. U.~O. 1993, Geophysical Research Letters, 20, 1763

\bibitem[{Hellinger \& Matsumoto(2000)}]{hellinger_new_2000}
Hellinger, P. \& Matsumoto, H. 2000, Journal of Geophysical Research: Space
  Physics, 105, 10519

\bibitem[{{Hollweg}(1974)}]{Hollweg74}
{Hollweg}, J.~V. 1974, \jgr, 79, 1539

\bibitem[{Horbury {et~al.}(2001)Horbury, Burgess, Fränz, \&
  Owen}]{horbury_three_2001}
Horbury, T.~S., Burgess, D., Fränz, M., \& Owen, C.~J. 2001, Geophysical
  Research Letters, 28, 677

\bibitem[{{Horbury} {et~al.}(2018){Horbury}, {Matteini}, \&
  {Stansby}}]{Horbury18}
{Horbury}, T.~S., {Matteini}, L., \& {Stansby}, D. 2018, \mnras, 478, 1980

\bibitem[{Horbury {et~al.}(2020)Horbury, Woolley, Laker, Matteini, Eastwood,
  Bale, Velli, Chandran, Phan, Raouafi, Goetz, Harvey, Pulupa, Klein, Dudok~de
  Wit, Kasper, Korreck, Case, Stevens, Whittlesey, Larson, MacDowall,
  Malaspina, \& Livi}]{horbury_sharp_2020}
Horbury, T.~S., Woolley, T., Laker, R., {et~al.} 2020, The Astrophysical
  Journal Supplement Series, 246, 45

\bibitem[{Hudson(1970)}]{hudson_discontinuities_1970}
Hudson, P. 1970, Planetary and Space Science, 18, 1611

\bibitem[{Jagarlamudi {et~al.}(2021)Jagarlamudi, Dudok~de Wit, Froment,
  Krasnoselskikh, Larosa, \& Bercic}]{Jagarlamudi_psp_whistlers}
Jagarlamudi, V.~K., Dudok~de Wit, T., Froment, C., {et~al.} 2021, under review,
  Astronomy and Astrophysics, this issue

\bibitem[{{Kasper} {et~al.}(2016){Kasper}, {Abiad}, {Austin}, {Balat-Pichelin},
  {Bale}, {Belcher}, {Berg}, {Bergner}, {Berthomier}, {Bookbinder}, {Brodu},
  {Caldwell}, {Case}, {Chand ran}, {Cheimets}, {Cirtain}, {Cranmer}, {Curtis},
  {Daigneau}, {Dalton}, {Dasgupta}, {DeTomaso}, {Diaz-Aguado}, {Djordjevic},
  {Donaskowski}, {Effinger}, {Florinski}, {Fox}, {Freeman}, {Gallagher},
  {Gary}, {Gauron}, {Gates}, {Goldstein}, {Golub}, {Gordon}, {Gurnee}, {Guth},
  {Halekas}, {Hatch}, {Heerikuisen}, {Ho}, {Hu}, {Johnson}, {Jordan},
  {Korreck}, {Larson}, {Lazarus}, {Li}, {Livi}, {Ludlam}, {Maksimovic},
  {McFadden}, {Marchant}, {Maruca}, {McComas}, {Messina}, {Mercer}, {Park},
  {Peddie}, {Pogorelov}, {Reinhart}, {Richardson}, {Robinson}, {Rosen},
  {Skoug}, {Slagle}, {Steinberg}, {Stevens}, {Szabo}, {Taylor}, {Tiu}, {Turin},
  {Velli}, {Webb}, {Whittlesey}, {Wright}, {Wu}, \& {Zank}}]{Kasper2016}
{Kasper}, J.~C., {Abiad}, R., {Austin}, G., {et~al.} 2016, \ssr, 204, 131

\bibitem[{{Kasper} {et~al.}(2019){Kasper}, {Bale}, {Belcher}, {Berthomier},
  {Case}, {Chandran}, {Curtis}, {Gallagher}, {Gary}, {Golub}, {Halekas}, {Ho},
  {Horbury}, {Hu}, {Huang}, {Klein}, {Korreck}, {Larson}, {Livi}, {Maruca},
  {Lavraud}, {Louarn}, {Maksimovic}, {Martinovic}, {McGinnis}, {Pogorelov},
  {Richardson}, {Skoug}, {Steinberg}, {Stevens}, {Szabo}, {Velli},
  {Whittlesey}, {Wright}, {Zank}, {MacDowall}, {McComas}, {McNutt}, {Pulupa},
  {Raouafi}, \& {Schwadron}}]{kasper19}
{Kasper}, J.~C., {Bale}, S.~D., {Belcher}, J.~W., {et~al.} 2019, \nat, 576, 228

\bibitem[{Kigure {et~al.}(2010)Kigure, Takahashi, Shibata, Yokoyama, \&
  Nozawa}]{kigure_generation_2010}
Kigure, H., Takahashi, K., Shibata, K., Yokoyama, T., \& Nozawa, S. 2010,
  Publications of the Astronomical Society of Japan, 62, 993

\bibitem[{Knetter(2004)}]{knetter_four-point_2004}
Knetter, T. 2004, Journal of Geophysical Research, 109, A06102

\bibitem[{Krasnoselskikh {et~al.}(2020)Krasnoselskikh, Larosa, Agapitov, Wit,
  Moncuquet, Mozer, Stevens, Bale, Bonnell, Froment, Goetz, Goodrich, Harvey,
  Kasper, MacDowall, Malaspina, Pulupa, Raouafi, Revillet, Velli, \&
  Wygant}]{krasnoselskikh_localized_2020}
Krasnoselskikh, V., Larosa, A., Agapitov, O., {et~al.} 2020, The Astrophysical
  Journal, 893, 93

\bibitem[{{Landi} {et~al.}(2005){Landi}, {Hellinger}, \& {Velli}}]{Landietal05}
{Landi}, S., {Hellinger}, P., \& {Velli}, M. 2005, in ESA Special Publication,
  Vol. 592, Solar Wind 11/SOHO 16, Connecting Sun and Heliosphere, ed.
  B.~{Fleck}, T.~H. {Zurbuchen}, \& H.~{Lacoste}, 785

\bibitem[{{Landi} {et~al.}(2006){Landi}, {Hellinger}, \& {Velli}}]{Landietal06}
{Landi}, S., {Hellinger}, P., \& {Velli}, M. 2006, \grl, 33, L14101

\bibitem[{Lepping \& Behannon(1980)}]{lepping_magnetic_1980}
Lepping, R.~P. \& Behannon, K.~W. 1980, Journal of Geophysical Research: Space
  Physics, 85, 4695

\bibitem[{Matteini(2015)}]{matteini_fire_2015}
Matteini, L. 2015, The Astrophysical Journal, 9

\bibitem[{{Moncuquet} {et~al.}(2020){Moncuquet}, {Meyer-Vernet}, {Issautier},
  {Pulupa}, {Bonnell}, {Bale}, {de Wit}, {Goetz}, {Griton}, {Harvey},
  {MacDowall}, {Maksimovic}, \& {Malaspina}}]{moncuquet20}
{Moncuquet}, M., {Meyer-Vernet}, N., {Issautier}, K., {et~al.} 2020, \apjs,
  246, 44

\bibitem[{{Mozer} {et~al.}(2020){Mozer}, {Agapitov}, {Bale}, {Bonnell}, {Case},
  {Chaston}, {Curtis}, {Wit}, {Goetz}, {Goodrich}, {Harvey}, {Kasper},
  {Korreck}, {Krasnoselskikh}, {Larson}, {Livi}, {MacDowall}, {Malaspina},
  {Pulupa}, {Stevens}, {Whittlesey}, \& {Wygant}}]{Mozer2020}
{Mozer}, F.~S., {Agapitov}, O.~V., {Bale}, S.~D., {et~al.} 2020, \apjs, 246, 68

\bibitem[{Neugebauer(2006)}]{neugebauer_comment_2006}
Neugebauer, M. 2006, Journal of Geophysical Research, 111, A04103

\bibitem[{Neugebauer {et~al.}(1984)Neugebauer, Clay, Goldstein, Tsurutani, \&
  Zwickl}]{neugebauer_reexamination_1984}
Neugebauer, M., Clay, D.~R., Goldstein, B.~E., Tsurutani, B.~T., \& Zwickl,
  R.~D. 1984, Journal of Geophysical Research, 89, 5395

\bibitem[{{Panasenco} {et~al.}(2020){Panasenco}, {Velli}, {D'Amicis}, {Shi},
  {R{\'e}ville}, {Bale}, {Badman}, {Kasper}, {Korreck}, {Bonnell}, {Wit},
  {Goetz}, {Harvey}, {MacDowall}, {Malaspina}, {Pulupa}, {Case}, {Larson},
  {Livi}, {Stevens}, \& {Whittlesey}}]{Panasenco20}
{Panasenco}, O., {Velli}, M., {D'Amicis}, R., {et~al.} 2020, \apjs, 246, 54

\bibitem[{Phan {et~al.}(2020)Phan, Bale, Eastwood, Lavraud, Drake, Oieroset,
  Shay, Pulupa, Stevens, MacDowall, Case, Larson, Kasper, Whittlesey, Szabo,
  Korreck, Bonnell, de~Wit, Goetz, Harvey, Horbury, Livi, Malaspina, Paulson,
  Raouafi, \& Velli}]{phan_parker_2020}
Phan, T.~D., Bale, S.~D., Eastwood, J.~P., {et~al.} 2020, The Astrophysical
  Journal Supplement Series, 246, 34

\bibitem[{Santolík {et~al.}(2003)Santolík, Parrot, \&
  Lefeuvre}]{santolik_singular_2003}
Santolík, O., Parrot, M., \& Lefeuvre, F. 2003, Radio Science, 38, n/a

\bibitem[{S\"oding {et~al.}(2001)S\"oding, Neubauer, Tsurutani, Ness, \&
  Lepping}]{soding_radial_2001}
S\"oding, A., Neubauer, F.~M., Tsurutani, B.~T., Ness, N.~F., \& Lepping, R.~P.
  2001, Annales Geophysicae, 19, 667

\bibitem[{{Sonnerup} \& {Scheible}(1998)}]{Sonnerup1998}
{Sonnerup}, B. U.~{\"O}. \& {Scheible}, M. 1998, ISSI Scientific Reports
  Series, 1, 185

\bibitem[{Squire {et~al.}(2020)Squire, Chandran, \& Meyrand}]{squire_situ_2020}
Squire, J., Chandran, B. D.~G., \& Meyrand, R. 2020, The Astrophysical Journal,
  891, L2

\bibitem[{Stansby {et~al.}(2019)Stansby, Perrone, Matteini, Horbury, \&
  Salem}]{stansby_alpha_2019}
Stansby, D., Perrone, D., Matteini, L., Horbury, T.~S., \& Salem, C.~S. 2019,
  Astronomy \& Astrophysics, 623, L2

\bibitem[{{Sterling} \& {Moore}(2020)}]{sterling20}
{Sterling}, A.~C. \& {Moore}, R.~L. 2020, The Astrophysical Journal Letters,
  896, L18

\bibitem[{{Stix}(1992)}]{stix1992}
{Stix}, T.~H. 1992, {Waves in plasmas} ({American Institute of Physics})

\bibitem[{Tenerani \& Velli(2018)}]{tenerani_nonlinear_2018}
Tenerani, A. \& Velli, M. 2018, The Astrophysical Journal, 867, L26

\bibitem[{Tsurutani {et~al.}(1994)Tsurutani, Ho, Smith, Neugebauer, Goldstein,
  Mok, Arballo, Balogh, Southwood, \& Feldman}]{tsurutani_relationship_1994}
Tsurutani, B.~T., Ho, C.~M., Smith, E.~J., {et~al.} 1994, Geophysical Research
  Letters, 21, 2267

\bibitem[{{Velli}(1993)}]{Velli93}
{Velli}, M. 1993, \aap, 270, 304

\bibitem[{Velli {et~al.}(1992)Velli, Grappin, \& Mangeney}]{velli92}
Velli, M., Grappin, R., \& Mangeney, A. 1992, AIP Conference Proceedings, 267,
  154

\bibitem[{Verniero {et~al.}(2020)Verniero, Larson, Livi, Rahmati, McManus,
  Pyakurel, Klein, Bowen, Bonnell, Alterman, Whittlesey, Malaspina, Bale,
  Kasper, Case, Goetz, Harvey, Korreck, MacDowall, Pulupa, Stevens, \&
  de~Wit}]{verniero_parker_2020}
Verniero, J.~L., Larson, D.~E., Livi, R., {et~al.} 2020, The Astrophysical
  Journal Supplement Series, 248, 5, arXiv: 2004.03009

\bibitem[{Verscharen \& Chandran(2013)}]{verscharen_dispersion_2013}
Verscharen, D. \& Chandran, B. D.~G. 2013, The Astrophysical Journal, 764, 88

\bibitem[{Whittlesey {et~al.}(2020)Whittlesey, Larson, Kasper, Halekas,
  Abatcha, Abiad, Berthomier, Case, Chen, Curtis, Dalton, Klein, Korreck, Livi,
  Ludlam, Marckwordt, Rahmati, Robinson, Slagle, Stevens, Tiu, \&
  Verniero}]{whittlesey_solar_2020}
Whittlesey, P.~L., Larson, D.~E., Kasper, J.~C., {et~al.} 2020, The
  Astrophysical Journal Supplement Series, 246, 74

\bibitem[{Woodham {et~al.}(2021)Woodham, Horbury, Matteini, Woolley, Laker,
  Bale, Nicolaou, Stawarz, Stansby, Hietala, Larson, Livi, Verniero, McManus,
  Kasper, Korreck, Raouaﬁ, Moncuquet, \& Pulupa}]{woodham_enhanced_nodate}
Woodham, L.~D., Horbury, T.~S., Matteini, L., {et~al.} 2021, 8

\bibitem[{Woolley {et~al.}(2020)Woolley, Matteini, Horbury, Bale, Woodham,
  Laker, Alterman, Bonnell, Case, Kasper, Klein, Martinović, \&
  Stevens}]{woolley_proton_2020}
Woolley, T., Matteini, L., Horbury, T.~S., {et~al.} 2020, Monthly Notices of
  the Royal Astronomical Society, 498, 5524

\bibitem[{{Yamauchi} {et~al.}(2004{\natexlab{a}}){Yamauchi}, {Moore}, {Suess},
  {Wang}, \& {Sakurai}}]{Yamab04}
{Yamauchi}, Y., {Moore}, R.~L., {Suess}, S.~T., {Wang}, H., \& {Sakurai}, T.
  2004{\natexlab{a}}, \apj, 605, 511

\bibitem[{{Yamauchi} {et~al.}(2004{\natexlab{b}}){Yamauchi}, {Suess},
  {Steinberg}, \& {Sakurai}}]{Yama04}
{Yamauchi}, Y., {Suess}, S.~T., {Steinberg}, J.~T., \& {Sakurai}, T.
  2004{\natexlab{b}}, Journal of Geophysical Research (Space Physics), 109,
  A03104

\end{thebibliography}

\end{document}